\documentclass[balance,upint,subscriptcorrection,varvw,mathalfa=cal=boondoxo,spanish,french,vietnamese,russian,greek,pdf-a,colorlinks]{asmeconf}

\usepackage{stfloats}
\usepackage{enumitem}
\usepackage{tabularx}
\usepackage{array}
\usepackage{booktabs}
\usepackage{mdframed}
\usepackage{graphicx}
\usepackage{colortbl}
\usepackage{enumitem}

\newcolumntype{L}{>{\small\bfseries\raggedright\arraybackslash}p{3cm}}

\newmdenv[
  topline=false,
  bottomline=false,
  rightline=false,
  linewidth=2pt,
  linecolor=blue,
  backgroundcolor=gray!20,
  leftmargin=10pt,
  rightmargin=10pt,
  innertopmargin=10pt,
  innerbottommargin=10pt
]{customquote}

\hypersetup{%
	pdfauthor={John H. Lienhard},									  
	pdftitle={ASME Conference Paper LaTeX Template},                  
	pdfkeywords={design inspiration, generative AI},
	pdfsubject = {Describes the asmeconf LaTeX template},			  
	pdflicenseurl={https://ctan.org/pkg/asmeconf},
}

\begin{document}


\ConfName{Proceedings of the ASME 2024\linebreak International Design Engineering Technical Conferences and \linebreak
Computers and Information in Engineering Conference}
\ConfAcronym{IDETC/CIE 2024-142124}
\ConfDate{August 25-28, 2024} 
\ConfCity{Washington, DC} 
\PaperNo{IDETC/CIE 2024-142124}


\title{Inspired by AI? \\A Novel Generative AI System To Assist Conceptual Automotive Design} 
 
%
%
%


\SetAuthors{
	Ye Wang\affil{1}\CorrespondingAuthor{ye.wang@autodesk.com}, 
	Nicole B. Damen\affil{1}, 
    Thomas Gale\affil{1},
	Voho Seo\affil{2},
    Hooman Shayani\affil{3}
	}

\SetAffiliation{1}{Autodesk Research}
\SetAffiliation{2}{Hyundai Motor Company}
\SetAffiliation{3}{Autodesk AI Lab}


\maketitle

\versionfootnote{Documentation for \texttt{asmeconf.cls}: Version~\versionno, \today.}




\begin{abstract}

Design inspiration is crucial for establishing the direction of a design as well as evoking feelings and conveying meanings during the conceptual design process. Many practice designers use text-based searches on platforms like Pinterest to gather image ideas, followed by sketching on paper or using digital tools to develop concepts. Emerging generative AI techniques, such as diffusion models, offer a promising avenue to streamline these processes by swiftly generating design concepts based on text and image inspiration inputs, subsequently using the AI generated design concepts as fresh sources of inspiration for further concept development. 
However, applying these generative AI techniques directly within a design context has challenges. Firstly, generative AI tools may exhibit a bias towards particular styles, resulting in a lack of diversity of design outputs. Secondly, these tools may struggle to grasp the nuanced meanings of texts or images in a design context. Lastly, the lack of integration with established design processes within design teams can result in fragmented use scenarios.
Focusing on these challenges, we conducted workshops, surveys, and data augmentation involving teams of experienced automotive designers to investigate their current practices in generating concepts inspired by texts and images, as well as their preferred interaction modes for generative AI systems to support the concept generation workflow. Finally, we developed a novel generative AI system based on diffusion models to assist conceptual automotive design.

\end{abstract}







\section{Introduction}

Conceptual design is a fluid and creative phase where designers collaborate to generate early designs that set the direction for development. Within this phase, designers work both collaboratively and independently. They are given the freedom to use a diverse set of tools, with some leaning towards traditional sketches while others prefer digital tools \cite{schenk2014inspiration}. Sketching plays a pivotal role in automotive concept development and is considered a way of visual thinking and communicating \cite{tovey2003sketching}. To foster collaboration, teams are often given the same context and concept keywords to guide their exploration. The role of the designers is to create novel designs that embody a strong concept, showcase uniqueness, and remain feasible and functional \cite{tovey2016designer}.

Both textual and image-based stimuli serve as frequent sources of inspiration in the concept design process. Textual stimuli excel at establishing a clear direction; for instance, a car design team might define ``bold'' and ``dynamic'' as key inspirations for the upcoming season's cars. In the automotive context, these keywords are not merely descriptive of a feeling, but have also come to refer to specific shapes as the field has evolved over the years \cite{costa2015temporal}. These keywords can then be communicated across different design teams—exterior, interior, and component—allowing them to work relatively autonomously while still achieving cohesive designs. 

On the other hand, images excel at evoking feelings and conveying meaning, with a closer connection to the physicality of design compared to text. From an image, designers can extract various visual elements such as colors, shapes, and textures, directly influencing their designs. Such elements can be used to evaluate the sales performance of previous models, and guide future developments \cite{jung2014discovery}. Moreover, designers can interpret textual concepts using imagery that goes beyond its car-related meaning; for instance, interpreting ``bold'' as reminiscent of the solidity of a concrete building or the stability of a vast river in nature.

Designers encounter various challenges when seeking inspiration and developing them into concepts. Firstly, the accessibility of online inspiration means that widely visible results are readily available to everyone. However, it can be time-consuming and idiosyncratic to manually curate or generate unique content \cite{herring2009getting}, and then translate them into design patterns that match the automotive context \cite{mirnig2016automotive}. Secondly, designers often employ specialized vocabulary that is not universally adopted for use by the vast majority of lay users on the internet \cite{saadi2023form}. For instance, terms like ``bold'' may carry nuanced meanings specific to an automotive design context, but may not manifest in the same way on public search databases. 

\begin{figure}[h]
    \centering
    \includegraphics[width=0.8\linewidth]{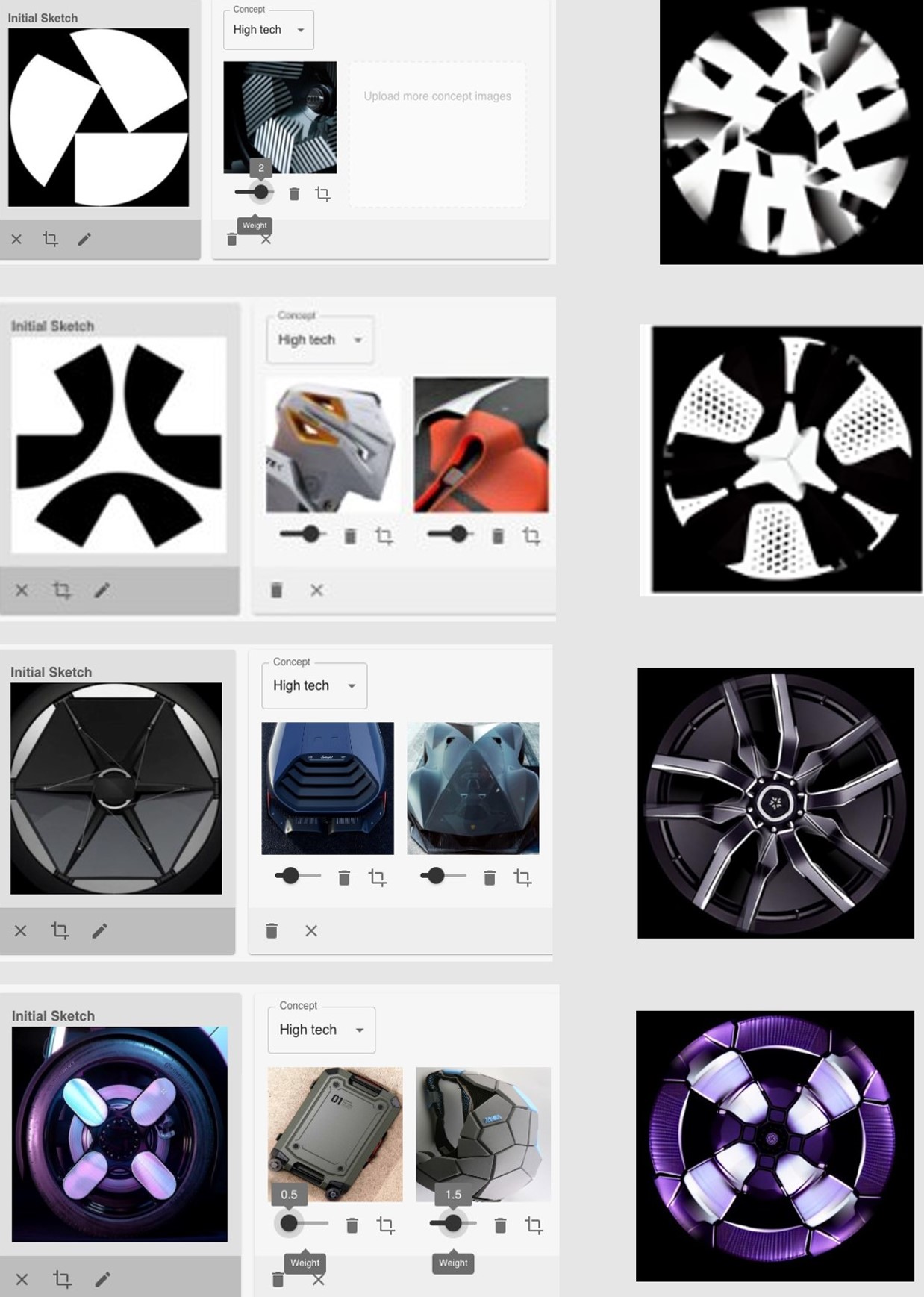}
    \caption{Example generation from designers. The right column are the designs generated by our AI system and the left are the text and image inspiration input from designers. }
    \label{fig:design_example}
\end{figure}

The emergence of generative AI techniques, especially diffusion models \cite{dhariwal2021diffusion}, holds promise for designers, with tools such as Midjourney \cite{Midjourney} and Stable Diffusion \cite{StableDiffusion} gaining traction among both professionals and amateurs for design creation. However, when employed within a design context, these tools may encounter several challenges. Firstly, generative AI tools may exhibit a tendency towards particular styles, leading to generated designs that resemble each other when provided with similar content \cite{cai2023designaid}. This limitation undermines the objective of incorporating diverse perspectives from different designers within the same team. Secondly, these tools may lack an understanding of the specific design contexts in which designers operate \cite{regenwetter2022deep}. Lastly, the absence of integration with established design processes results in more isolated use cases, limiting their overall effectiveness \cite{saadi2023generative}.

To use generative AI for design inspiration, novelty and diversity play crucial roles. As expressed by one designer in an interview conducted for this work, the appeal of generative AI lies in its ability to offer fresh and unique perspectives, diverging from the designer's habitual tendencies. Moreover, design teams possess their own distinct processes; hence, seamless integration of generative AI within the conceptual design workflow would facilitate its adoption among designers.

Thus, the goal of this research is to create a novel generative AI system to support designers to swiftly generate concept designs by integrating their individual and team inspiration. 
With this goal in mind, we focused on teams of car designers operating within a single company. Our approach involved conducting user studies, hosting workshops, collecting data, and prototyping and developing a generative AI system to address the following three research questions: 
\begin{enumerate}
    \item How are designers currently integrating text and image inspiration into their conceptual design practices? (Section \ref{sec:design_inspiration_studies} \nameref{sec:design_inspiration_studies})
    \item What interaction modes do designers prefer for generative AI systems to develop concepts from design inspiration? (Section \ref{sec:interaction_mode} \nameref{sec:interaction_mode})
    \item What are the specifications for building a generative AI system to facilitate the interaction modes desired by designers? (Section \ref{sec:ai_system} \nameref{sec:ai_system})
\end{enumerate}

Design encompasses both personal expression and collaborative effort. To effectively integrate new generative AI systems into designers' workflows, the technology must embody human creativity while remaining grounded in human-centered design processes. This endeavor aims to investigate the potential characteristics of such systems. Spanning from initial interviews to the final development of the system and user testing, this project spanned a year, coinciding with numerous AI advancements. The rapid evolution of such potent technology will undoubtedly impact the design field, presenting both opportunities and challenges. Studies like this endeavor serve to introduce AI technologies that empower designers while prompting critical questions about the future of design.

\section{Background}
\subsection{Design Inspiration}

Inspiration plays a crucial role in conceptual design. It serves as both an analogy and a guide, facilitating exploration across a vast design space \cite{fu2013meaning,linsey2008increasing,jiang2022data}. Additionally, inspiration helps designers in overcoming fixation and generating novel ideas \cite{vasconcelos2016inspiration, crilly2015fixation,vasconcelos2017inspiration}, defining design directions \cite{gonccalves2016inspiration}, and communicating design concepts and contexts \cite{eckert2000sources}. Inspiration manifests in various forms, including drawings, texts, images, 3D designs, and physical products such as teardowns \cite{schenk2014inspiration,goucher2019crowdsourcing, jonson2005design,toh2014impact}, and can be acquired intentionally through search or naturally through experience \cite{herring2009getting,gonccalves2011around}. These forms of inspiration can be encountered physically, digitally, and virtually \cite{hu2020mitigating,herring2009getting,goucher2020adaptive}, facilitating the creation of new concepts.

This work concentrates on text and image inspiration for several reasons. Firstly, these forms are easily and affordably accessible in large quantities and varieties via the internet, and are commonly utilized by designers. Secondly, text and images are typically integrated into design processes, serving as tools for establishing design directions and communicating ideas. Lastly, many advanced generative AI models focus on text and image inputs and outputs, benefiting from the abundance of high-quality content available.

This work emphasizes the digital experience of inspiration among designers, aligning with the processes observed in the automotive design teams under study. Designers primarily source and generate inspiration digitally, with AI-generated designs later serving as image inspiration for further digital concept exploration.

The studies center on expert designers' use of inspiration and concept generation. It is important to acknowledge that amateur users and professional designers may seek different interaction modes with generative AI. Professional designers with greater expertise typically engage in more complex design processes \cite{atman2019design}. Therefore, understanding and integrating new generative AI tools into professional designers' workflows will be crucial for ensuring their effectiveness in conceptual design.

\subsection{Generative AI for Image Generation}
\label{sec:generative_ai}
Variational Auto-Encoders (VAEs) \cite{kingma2013auto} and Generative Adversarial Networks (GANs) \cite{goodfellow2020generative} have been two classical deep generative models that were used for generating images with relative success. Recently, Denoising Diffusion Probabilistic Models (DDPMs) \cite{dhariwal2021diffusion} have shown even greater success in generating high-quality and diverse images with better training stability. These models can learn to predict the reverse of a diffusion process (adding Gaussian noise) applied to the data, gradually denoising an image from noise to clean data points similar to the images in the training dataset \cite{ho2020denoising}. However, the need to iterate through many denoising steps makes the original DDPMs slower than GANs and VAEs.

To tackle this problem, Latent Diffusion Models (LDMs) such as Stable Diffusion \cite{rombach2022high} first train a high-quality Auto-Encoder that learns a latent representation for the images and then perform the denoising diffusion process on that smaller latent tensor. This reduces the size and computational cost of these models substantially. These models are usually conditioned on a textual prompt that describes the image semantically.

To provide users with fine-grained control over the general composition, shape, and position of the generated images, new models such as ControlNet \cite{zhang2023adding} provide additional conditional inputs that influence the generation. Models that can map text and images into a common latent space, such as Kandinsky \cite{razzhigaev2023kandinsky}, allow these modalities to be mixed and collectively used to control the generated images. Versatile Diffusion \cite{xu2023versatile} supports many different use-cases, including using multiple images and textual prompts as conditional inputs using cross-attention mechanisms. Particularly, Versatile Diffusion uses a global signal over the input images that can be used in the cross-attention, which causes the input image to influence the semantics of the generated image more than its style and local features.

\subsection{AI Usage in Conceptual Design}
\label{sec:AItools}

Various design tools have been developed previously for searching and exploring inspiration \cite{kwon2023understanding,kwon2022enabling}, as well as generating design concepts. Many of these tools leverage large language models (LLMs) to aid designers in brainstorming, exploring user and engineering requirements, searching for design examples, and synthesizing design ideas \cite{wang2023task,ma2023conceptual,zhu2023generative,zhu2023biologically}.

In this study, we focus on image concepts, where designers utilize both text and images to inspire car wheel designs. As AI expands its capabilities it increases its potential for use in the design process. For example, AI support text-based ideation has been applied to product design \cite{filippi2023measuring}, and there are existing tools built on generative AI for image generation \cite{brisco2023exploring,cai2023designaid}. For example, progressive sketching has been used to stimulate design creativity \cite{cluzel2012using}.

However, these tools do not address certain outstanding challenges. Firstly, these generative AI tools may exhibit biases towards particular styles, resulting in a lack of diversity in design outputs. Secondly, they may struggle to interpret the nuanced meanings of texts or images within a design context. Lastly, the lack of integration with established design processes within design teams can lead to fragmented usage scenarios.



\section{Method}
\label{sec:method}
\begin{figure}[h]
    \centering
    \includegraphics[width=0.75\linewidth]{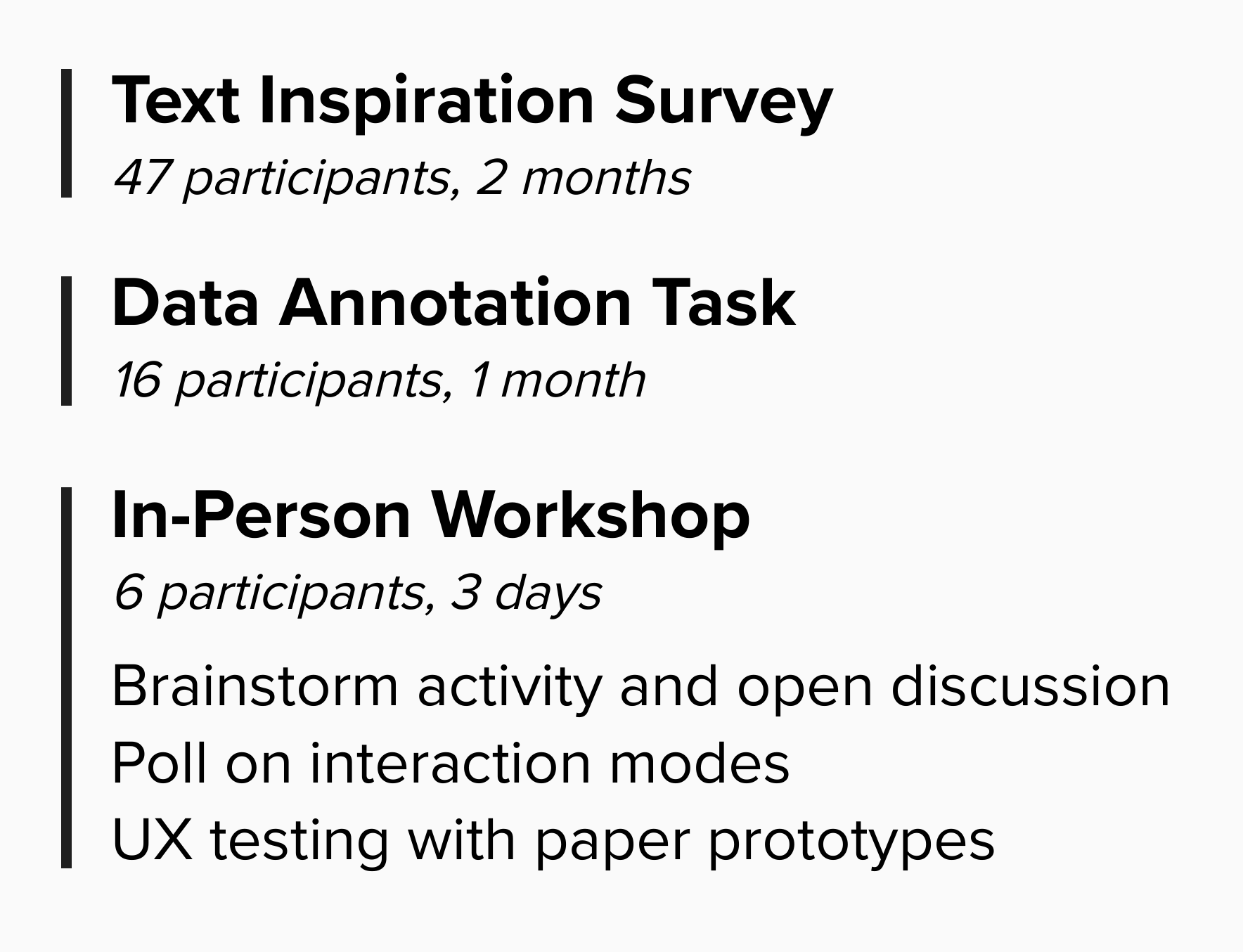}
    \caption{The methods used in the paper are presented in chronological order. Participants were recruited separately for each of the three activities.}
    \label{fig:method}
\end{figure}

In order to address the three challenges mentioned earlier and develop an innovative generative AI system, data was gathered from diverse sources. As illustrated in Figure \ref{fig:method}, first, designers were asked to fill in a survey about textual inspiration (Section \ref{sec:survey}), then complete a Data Augmentation Task (Section \ref{sec:data_collection}), and lastly a three-day in-person workshop (Section \ref{sec:workshop}). These methods were selected to provide both qualitative insights and quantitative coverage, ensuring that the conclusions drawn from in-depth analysis are representative of a broader spectrum of designers. The following sections detail these methods (see \ref{fig:method}).

\subsection{Text Inspiration Survey}
\label{sec:survey}
We conducted a survey to explore how automotive designers incorporate text inspiration into their design processes. A total of 47 designers participated in the survey, boasting an average of 12 years of design experience in various design specialties such as interior, exterior, component, design modeling, and future design. Among them, 6 designers have five years of experience or less, 15 have six to ten years, and 26 have more than ten years of experience. 

From the survey, we focused on two sets of questions for analysis in this paper. The first set inquired about the usefulness of keywords in their design process: ``How useful are design keywords when you generate design ideas or communicate your designs to others?'' and ``What about keywords makes them useful or not useful to you? ''. The second set asked participants to identify the most commonly used keywords for their design context: ``Given the following list of keywords, please select the top 10 keywords that you used most often in the past year(s)?'' and ``What do you think is the reason you used those keywords most often?''. Section \ref{sec:keyword_direction} \nameref{sec:keyword_direction} and Section \ref{sec:keyword_meaning} \nameref{sec:keyword_meaning} present the findings derived from these questions.

\subsection{Data Augmentation Task}
\label{sec:data_collection}
\begin{figure}[h]
    \centering
    \includegraphics[width=0.75\linewidth]{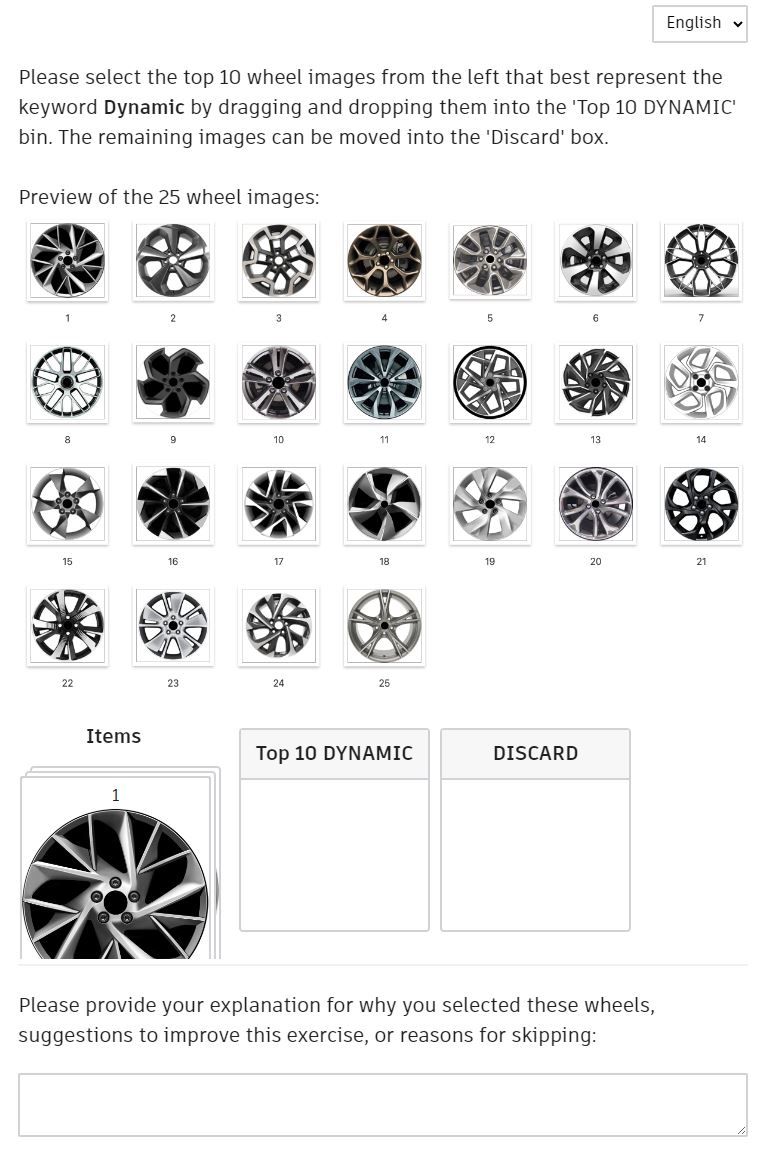}
    \caption{Interface of the data augmentation task.}
    \label{fig:annotation_survey}
\end{figure}

To establish a foundational understanding of the relationship between inspiration keywords and car wheel designs, we conducted an annotation task with 16 automotive designers. These designers possess an average of 12 years of experience, with 4 having five years of experience or less, 3 with six to ten years, and 9 with over 10 years of experience. The task, lasting approximately 30 minutes, involved presenting designers with 25 car wheels and asking them to select the top 10 wheels that best represent a given keyword, such as ``dynamic''. These 25 car wheels corresponded to ten keywords selected from those most commonly used, as identified in the text inspiration survey. An example task for the keyword ``dynamic'' is illustrated in Figure \ref{fig:annotation_survey}. 

The exemplary wheels were determined based on the frequency of selection by designers, with the top $5\%$ most chosen wheels selected to represent each keyword. The number of exemplary wheel designs varied for each keyword, ranging from 4 to 11. These examples were subsequently utilized in the AI model (Section \ref{sec:ai_model_architecture}) to establish a foundational understanding of these keywords within the design context.

\subsection{In-Person Workshop}
\label{sec:workshop}
We organized a three-day workshop with six automotive designers to explore their approach to drawing inspiration in conceptual design and their preferred interaction modes with generative AI technologies. Although they worked at the same company, these designers were not familiar with each other prior to the workshop as they worked in different teams. Their expertise spans interior, exterior, and component car design, as well as digital modeling, product planning, and future design. 

The designers were split evenly between those with 2 to 5 years of design experience and those with over 10 years of experience. Throughout the in-person workshop, we guided them through the following three activities using a combination of digital (Mural) and analog (Pen and paper, paper prototype) tools.

\subsubsection{Brainstorm activity and open discussion:} Designers utilized Mural, a collaborative virtual canvas where people can brainstorm and organize ideas using sticky notes, images, and drawings. They were tasked with generating image inspiration for eight keyword prompts, such as ``bold'', and elaborating on how specific designs were associated with each keyword. This exercise provided insights into the connection between text inspiration and designs in the minds of the designers. The findings from this activity are detailed in Section \ref{sec:embody_keywords} \nameref{sec:embody_keywords}. Additionally, we prompted designers to provide examples illustrating the distinction between being ``inspired'' and ``influenced'', recognizing concerns regarding inadvertently replicating designs through generative AI technologies. Section \ref{sec:inspired_not_copied} \nameref{sec:inspired_not_copied} addresses this aspect.

\subsubsection{Poll on interaction modes:} Designers were asked to answer seven questions regarding their preferred interaction modes when using generative AI for design inspiration and concept development. These questions were delivered via Mentimeter \cite{Mentimeter}. Each participant used their own mobile device to provide their answer. Once all answers were collected, the results were displayed, and designers engaged in group discussions. Table \ref{tab:interaction_mode_questions} shows the list of questions. The findings informed our development of our system. Section \ref{sec:interaction_mode} \nameref{sec:interaction_mode} presents and discusses these findings.

\subsubsection{User experience testing with paper prototypes:}  
\begin{figure}[h]
    \centering
    \includegraphics[width=0.8\linewidth]{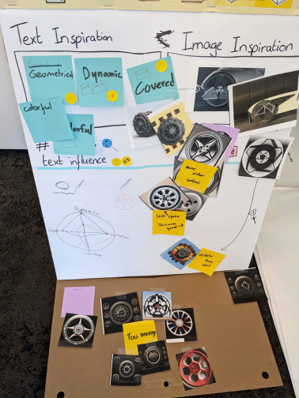}
    \caption{User Testing with Paper Prototypes}
    \label{fig:paper_prototype}
\end{figure}

We conducted tests with paper prototypes of an AI interface to investigate designers' preferred interaction modes for integrating inspiration into concept development. 
Six designers were equally divided into two groups. Each group interacted with the paper prototype, assuming the role of the user inputting inspiration, while a facilitator acted as the AI generating design outputs. Designers were encouraged to think aloud and engage in interactions beyond the constraints of the paper prototype. For example, designers could directly point at a generated design, and tell the facilitator the actions they wanted to perform on that design. An example of designer feedback on the paper prototype is shown in Figure \ref{fig:paper_prototype}. 


For input, each group received 50 pre-printed internet images. Designers could also sketch on sticky notes to contribute inspiration. The facilitators were provided with 50 pre-printed images of car wheels and were permitted to modify them to ensure output flexibility. The findings from this prototype testing are detailed and discussed in Section \ref{sec:image_inspiration} \nameref{sec:image_inspiration} and Section \ref{sec:interaction_mode} \nameref{sec:interaction_mode}.   

\begin{table}[h]
\renewcommand{\arraystretch}{1.5}
    \centering
    \begin{tabular}{|c|p{0.38\textwidth}|}
        \hline
         Q1 & What types of inspiration image do you want?\\
         \hline
         Q2 & What are the top text input option(s) you would prefer most? \\
         \hline
         Q3 &  How many style concepts would you like to express in your textual prompt?\\
         \hline
         Q4 &  How many images and drawings would you like to input to influence the AI generation?\\
         \hline
         Q5 & What do you expect the 2D output to be like for it to be useful for you? \\
         \hline
         Q6 & Rank the following six designs on novelty and usefulness.\\
         \hline
         Q7 & When do you want the ``generation'' to happen? \\
         \hline
         Q8 & Rank the design organization features from most to least useful. \\
         \hline
    \end{tabular}
    \caption{Workshop Poll}
    \label{tab:interaction_mode_questions}
\end{table}

\section{Findings on Design Inspiration}
\label{sec:design_inspiration_studies}

In conceptual design, inspiration plays a crucial role in establishing and communicating design directions, as well as evoking feelings and conveying meanings. In this section, we present and analyze the findings from both survey and the workshop to explore the use of text and image inspiration in the process of car design such as the distinction between ``inspired'' versus ``influenced''. 

\subsection{Text Inspiration}
\label{sec:text_inspiration}
The design of complex products such as cars typically involves the collaborative efforts of multiple teams. To facilitate autonomous work while ensuring a cohesive design concept, designers across various teams are often provided with a set of \textit{keywords} to establish the direction. Text serves as a valuable tool for conveying ideas without imposing the same level of constraints as images \cite{costa2015temporal, tovey2016designer}. Our goal in this study is threefold: firstly, to comprehend how designers integrate text inspiration into conceptual design; secondly, to identify specific texts that hold greater significance within a particular design context, in our case, a car design company; and finally, to examine the connection between these texts and the final designs.

\subsubsection{Using keywords to define design directions}
\label{sec:keyword_direction}
To better understand how designers employ keywords and evaluate how appropriate they might be in an AI tool, we included a survey question regarding the usefulness of design keywords: ``How useful are design keywords when you generate design ideas or communicate your designs to others?''. Designers were provided with four response options: Not at all useful, somewhat useful, very useful, and extremely useful. Out of $46$ designers who participated in the survey, $16$ found keywords extremely useful, $18$ very useful, and $12$ somewhat useful. No respondent rated keywords as not useful. In sum, designers appreciate keywords for their ability to simultaneously express style, form, and function while leaving the exact expression open for creative interpretation

Designers who indicated that keywords are very useful or extremely useful find them to be a quick and easy way to express and evaluate a design, guiding its direction. One designer noted, ``the right keywords give the right intention to the listener and help to imagine the concept much better in a very short time.''. Another designer expressed, ``They help me quickly understand, visualize, and analyze the essence of the project, form, or topic''. Additionally, designers appreciate that keywords help them maintain focus: ``I can continue thinking about my design without losing direction, and it is easy to explain to others what part of the design I want to emphasize''. 

However, some designers view keywords as somewhat useful due to their limitations in conveying precise details such as feelings and numerical values. One designer remarked, ``Quick and easy communication, however, the accuracy seems to be poor''. Another designer stated, ``There is a limit to capturing all the emotional parts with keywords''. One designer commented, ``design is a synthesis of images, keywords, and other designer's living environment (cultural) attitudes.''.  Another designer highlighted, ``Keywords are important, but it is better to focus on the message that the images should convey''.

\subsubsection{Shared vocabularies within a design context}
In the same survey, we prompted designers to reflect on the top ten keywords they used most often in projects in the past year. Designers were provided with $37$ keywords, such as ``boxy'', ``sharp'' and ``agile'', curated from their design documents.  $39$ designers answered this question, and on average each designer selected at least nine keywords. This question aimed to investigate whether designers within the same design context, in this case, working in the same car design company, tend to develop similar design vocabularies. The findings of this question has implications for the design of the AI tool, as it may affect how similar users may interact differently with the same tool. 

The results reveal that some keywords were notably recurrent among designers. The top three keywords, ``modern'', ``dynamic'' and ``bold'', were selected by $29$, $27$ and $27$ designers out of $39$ in total. All $39$ designers chose at least one of these top three keywords. From there, the consensus quickly drops - the ninth top selected keyword was selected by 14 designers, half less than the top selected keyword. This indicates that designers within this design context have indeed established a shared vocabulary for communicating their designs, and they possess a mutual understanding of the meanings within this vocabulary. 

\label{sec:keyword_meaning}
\subsubsection{Translating keywords into designs}
\label{sec:embody_keywords}

\begin{figure}[h]
    \centering
    \includegraphics[width=0.5\linewidth]{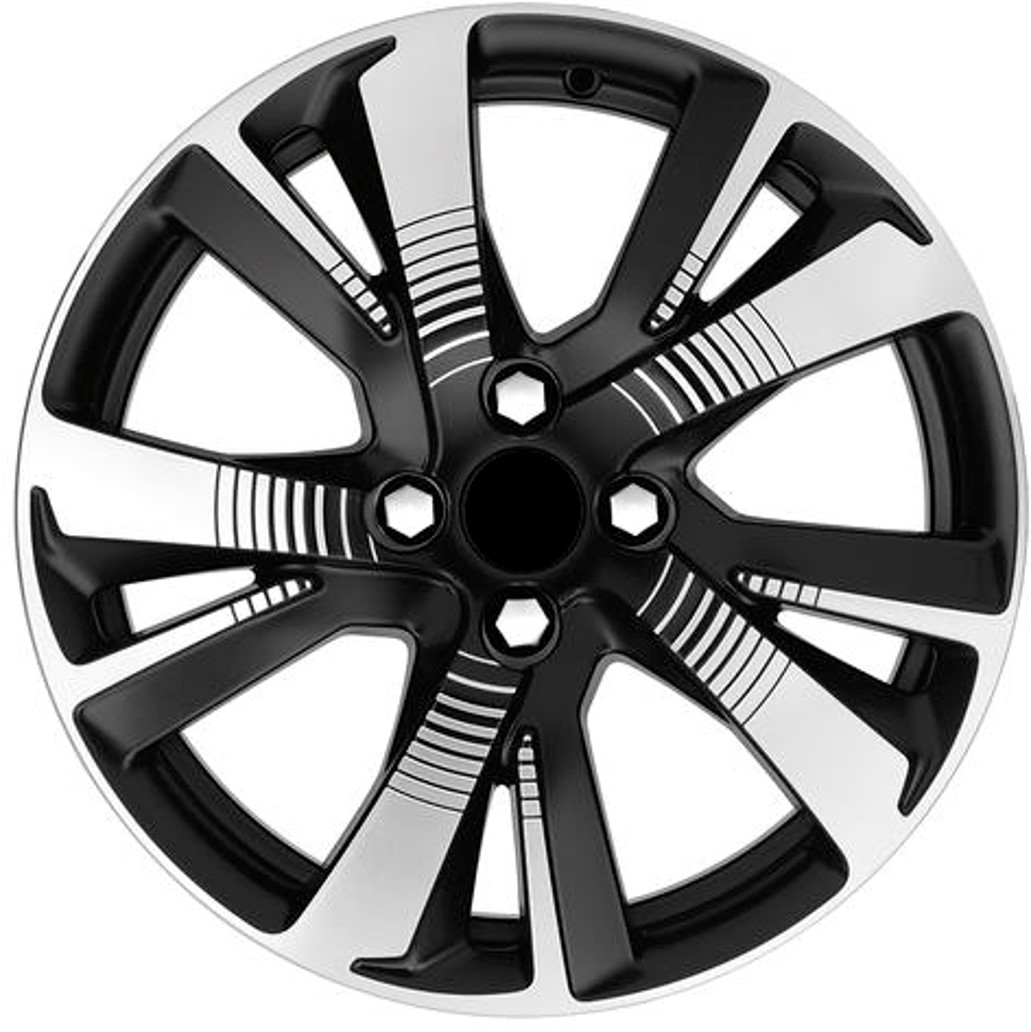}
    \caption{A dynamic wheel.}
    \label{fig:dynamic_wheel}
\end{figure}

During the workshop, designers were tasked with generating keywords to describe a particular design. For the wheel example in Figure \ref{fig:dynamic_wheel}, designers unanimously used ``dynamic'' to define the design. Subsequently, we asked them to elaborate on why they perceived the design as ``dynamic''. Here are some insights shared by the designers:
\begin{itemize}
    \item ``This wheel is dynamic since the gaps have different sizes. That creates a feeling of \textit{motion}. '' 
    \item ``The \textit{varied} radial patterns engraved at the center of the wheel creates a sense of \textit{speed}. ''
    \item ``The angled cutout from the tip of the smaller gaps indicate \textit{directions}, which give a feeling of \textit{speed}.'' 
\end{itemize}

Through this activity, it became evident that designers interpret the keywords in diverse ways, and translate these interpretations into visual characteristics within their designs. Notably, none of these explanations and translations were formally documented in the design process. This lack of documentation poses challenges for creating AI systems capable of understanding text inspiration in design contexts. Consequently, additional design data annotation is necessary to facilitate AI comprehension of the associations between keywords and designs. Section \ref{sec:data_collection} \nameref{sec:data_collection} explains the annotation process.

\subsection{Image Inspiration}
\label{sec:image_inspiration}
During the workshop, designers were tasked with searching for images that would inspire their design concept. It was observed that designers use image inspiration to convey feelings, meanings, and incorporate visual elements, such as shapes, textures, and colors into their designs. 
This section explores how designers leverage image inspiration and demonstrates how it compliments text inspiration in concept development. 

\subsubsection{Convey feelings with image inspiration}
\label{sec:convey_feelings}
\begin{figure}[h]
    \centering
    \includegraphics[width=\linewidth]{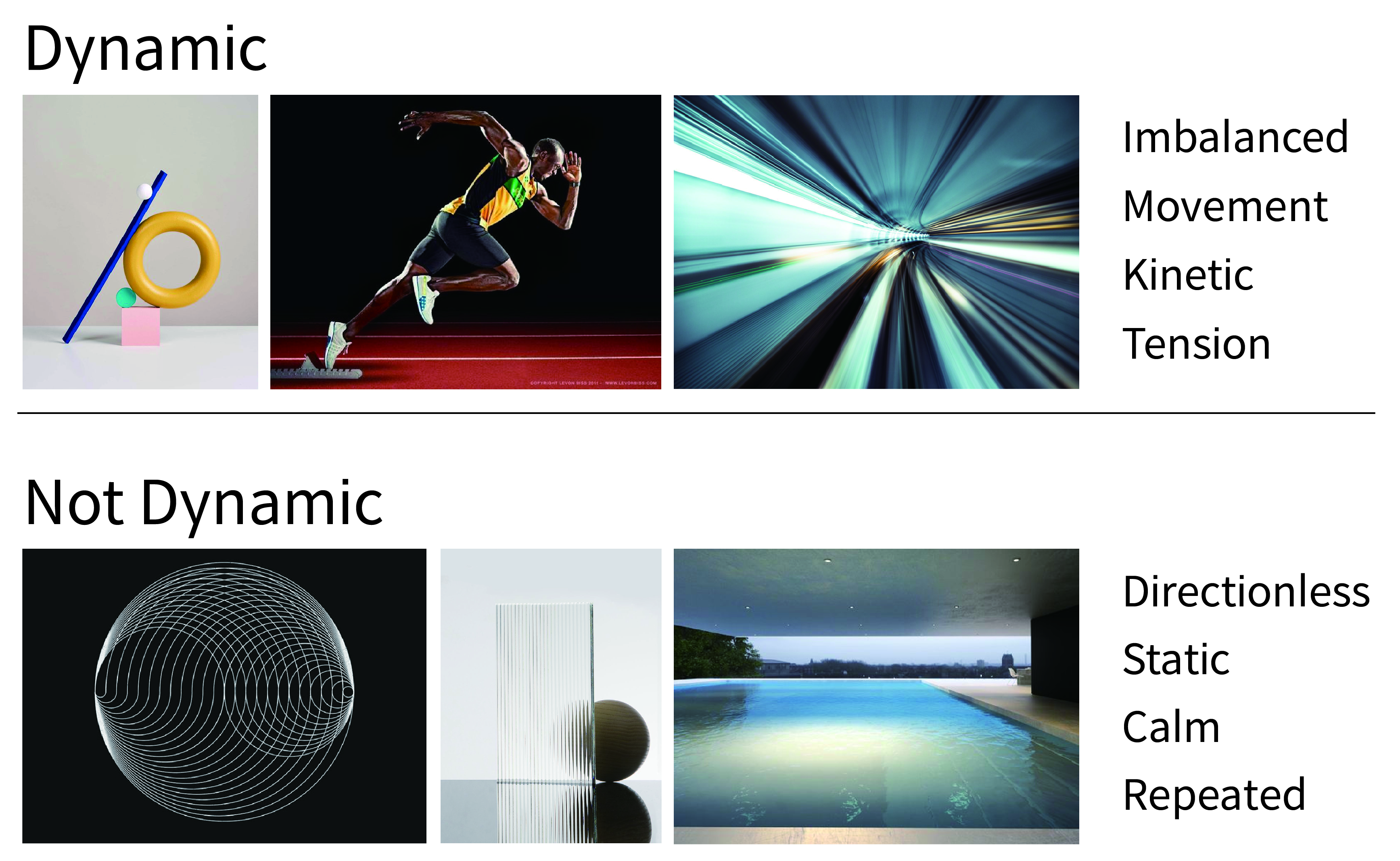}
    \caption{Text and image inspiration from designers for ``dynamic'' and ``not dynamic''.}
    \label{fig:dynamic_inspiration}
\end{figure}

In response to Question 1 of the workshop poll, ``what types of inspiration image do you want?'', two out of six designers emphasized the importance of images to communicate feelings. One designers suggested ``an instance image that represents a certain feeling,'' while another simply mentioned ``mood''. During user interaction testing with paper prototypes,  designers consistently chose images that convey feelings rather focusing solely on shapes. For example, one designer selected an image of a gun to communicate the feeling of ``agressive'', stating, ``it is the symbolism behind the object, not its literal shape, that matter''. 

Figure \ref{fig:dynamic_inspiration} shows an example where a designer uses an image of delicately stacked geometric objects to convey a dynamic concept. The designer explains that the imbalance embodied in the image can inspire designers to create designs that feel ``dynamic''.

\subsubsection{Extract visual elements from image inspiration} 
\label{sec:extract_visual_elements}
When tasked with finding image inspiration for ``not dynamic'' during the workshop, designers selected images depicting straight lines, symmetry and repeated as in Figure \ref{fig:dynamic_inspiration}. Designers can extract the shapes and patterns directly from these inspiration images, integrating these visual elements into their designs. In response to Question 1 of the workshop poll, designers mentioned various types of image inspiration, including ``simple thumbnail styled basic geometry, not so much detail'', ``all industrial product image'', and ``color''. Furthermore, during the paper prototype, designers incorporated image inspiration similar to those shown in Figure \ref{fig:paper_prototype}, highlighting specific shapes to inspire the generation of new concepts.

\subsubsection{Inspired v.s. influenced}
\label{sec:inspired_not_copied}

\begin{figure*}[ht]
    \centering
    \includegraphics[width=0.95\linewidth]{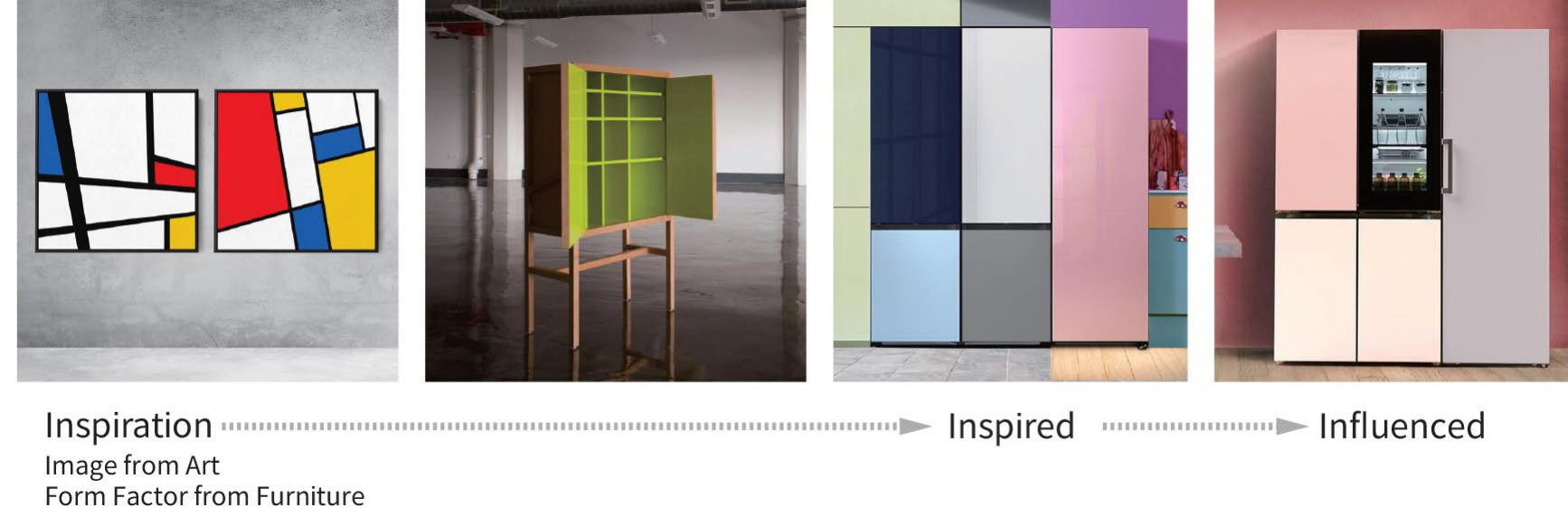}
    \caption{Example of inspired versus influenced. Extracting feelings or visual elements from designs originating in different fields—such as art and furniture—and translating and incorporating them into their own design context—such as home appliances—is considered as ``inspired''. However, directly using elements from a similar design field and applying them in their own design would be seen as ``influenced''. }
    \label{fig:inspired_copied}
\end{figure*}

During the workshop, we asked designers to illustrate examples of designs they considered as ``inspired'' versus ``influenced''. As depicted in Figure \ref{fig:inspired_copied} by the designers, they explained that extracting feelings or visual elements from designs originating in different fields—such as nature, art, and furniture—and translating and incorporating them into their own design context—such as home appliances—is considered as ``inspired''. However, directly using elements from a similar design field and applying them in their own design would be seen as ``influenced''. 

In response to Question 1 of the workshop poll regarding the types of inspiration image designers want, three out of six designers mentioned designs unrelated to cars. We observed similar behavior during user experience testing with paper prototypes, where designers frequently used image related to art and product design. In the rare cases where they used car-related images, they drew simple lines with white markers, expressing a preference for AI to only use these lines.

Novelty is crucial in conceptual design. Understanding designers' expectations on the types of text and image inspiration they wish to use, as well as how they anticipate the inspiration being incorporated into the generation of new concepts, is critical for designing new AI assistance for this design phase. In the following section, we will explore designers' preferred interaction modes with AI systems.  

\section{Findings on Interaction Modes}
\label{sec:interaction_mode}
This section will present and discuss the findings from both the workshop poll and the user experience testing with paper prototypes, focusing on designers' preferred interaction with AI. The findings are organized into three subsections: designers' input, design output, and integration into the design process. 

\subsection{Inspiration Inputs from Designers}
\subsubsection{Keywords for concept definition}
\label{sec:keyword_concept_definition}
In Question 2 in the workshop poll, ``What are the top text input option(s) you would prefer most?'', designers were presented with six options for multiple choices, along with an ``other'' option to accommodate additional ideas. Table \ref{tab:top_text_input} shows these text input options alongside the corresponding response counts. Notably, none of the designers selected the ``other'' option, indicating that the provided choices adequately captured their preferences. All six designers expressed the importance of keywords in setting design directions, prioritizing them over detailed shape instructions and requirements. They emphasized that during the design review and refinement process, they typically provide feedback to enhance detailed shapes or propose additional variations. Integrating such inputs into an AI tool could facilitate these processes and stimulate the generation of new ideas for designers.

\begin{table}[h]
\renewcommand{\arraystretch}{1.5}
    \centering
    \begin{tabular}{|p{0.32\textwidth}|c|}
        \hline
         \textbf{Text Input Option} & \textbf{Responses}\\
         \hline
         keywords for styles (e.g. bold, modern) & 6 \\
         \hline
         detailed shape instruction (e.g. enhanced character lines, sharp angles) &  5\\
         \hline
         requirements (e.g. four spokes) &  4\\
         \hline
         design scenario (e.g. a car for American market) & 3 \\
         \hline
         overall vehicle style description (e.g. an off road car) & 2\\
         \hline
         image style output (e.g. realistic rendering, pencil sketch) & 1 \\
         \hline
         Other & 0 \\
         \hline
    \end{tabular}
    \caption{Top text input option(s) designers prefer}
    \label{tab:top_text_input}
\end{table}

\subsubsection{Quantity of inspiration input} 
\label{sec:quantity_inspiration}
In Question 3 and 4 during the workshop, designers were asked about their preferred number of text and image inspiration inputs. For text inspiration, designers preferred two to four inputs to maintain a focused design direction. As one designer articulated, ``I'll input 3 to 4 concepts. This is because too little information is insufficient, and too much would make design direction unclear.'' Additionally, two designers mentioned the hierarchical refinement of design concepts with text inputs. One designer suggested, ``At least 2 to 3 main keywords for main stream and 2 to 3 detailed keywords to explain the theme''. Another designer echoed this viewpoint, proposing, ``At least two. Designer will expect very few keywords to narrow it (the concept) down. For example, dynamic can be agile instead of bold.''  

Regarding image inspiration, designers exhibited more variability in their preferences. Half of the six designers indicated a preference for one to three images. One designer explained, ``1 to 2 images contain feelings. And 2 to 3 image instances of the application to make the direction clear.'' Another designer specified, ``one for rendering style, one for design sketch for the theme. And one for refined details.'' Conversely, two designers desired more than five images but shared similar concerns about clarify. As one explained, ``I think I will put in about five. Too much will make the direction rather unclear. What matters is the consistency of the reference image.'' 

One designer didn't specify a numerical value but instead described the desired interaction with AI generation. ``Be able to add or remove the reference images after getting the output to see the updates.'' This aligns observations from the user experience testing with paper prototypes in Section \ref{sec:regenerate_with_feedback} \nameref{sec:regenerate_with_feedback}. 

In summary, designers prefer minimal text and image inspiration to establish a clear design direction, with the addition of consistent image inspiration to explore visual ideas. The quantity of images also depends on the AI system's ability to interpret designers' intentions, which may result in a hierarchical approach to inspiration. 

\subsubsection{Hierarchical approach to inspiration} 
\label{sec:hiearchical_approach}
\begin{figure}[ht]
    \centering
    \includegraphics[width=\linewidth]{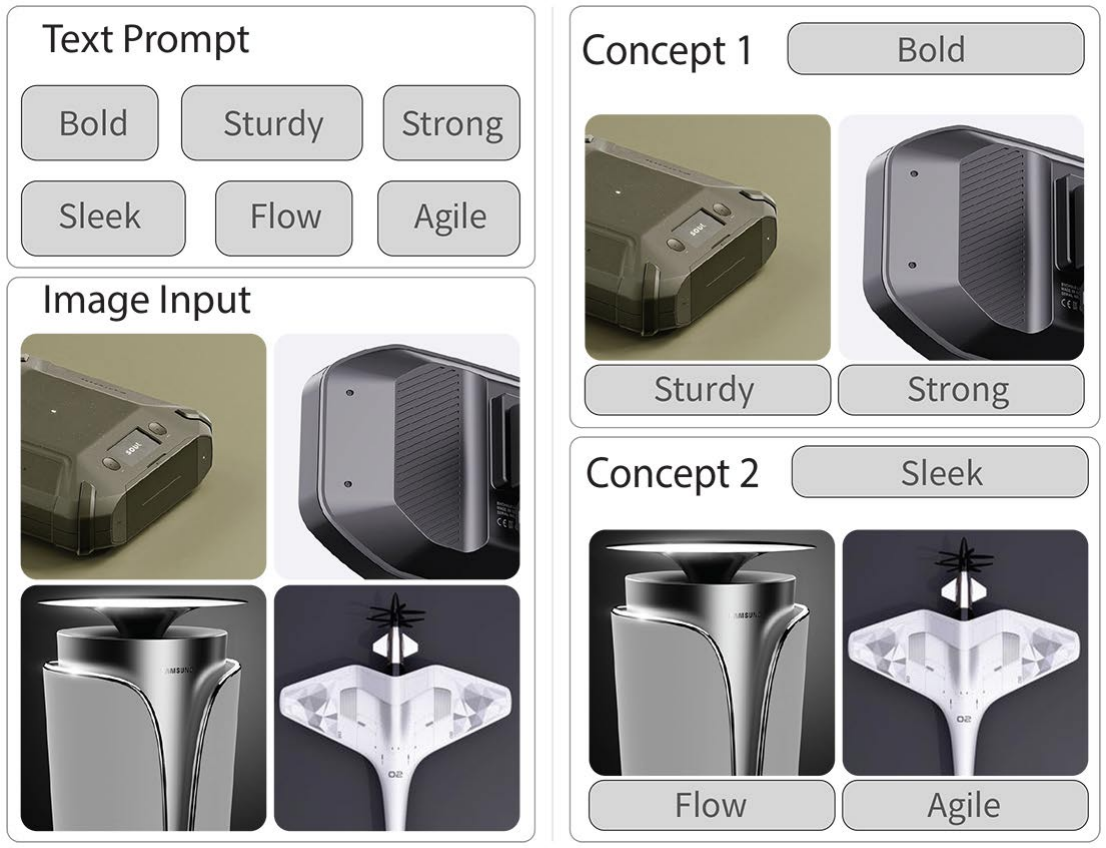}
    \caption{Contrast between hierarchical (right) and non-hierarchical approaches to inspiration (left). Instead of grouping prompts by type, such as text and images, user testing revealed that designers prefer to group them by concept. Text sets the direction for the concept, while images help guide the visual ideas.}
    \label{fig:hiearchy}
\end{figure}

Rather than categorizing input inspiration into traditional types like image, text, and sketch, designers adopt a hierarchical perspective. They use keywords, such as ``bold'' and ``dynamic'', to establish design directions. As designers develop these concepts, they incorporate detailed text and image inspiration and explore the generative AI output. Figure \ref{fig:hiearchy} illustrates this approach. This preference is evident in the user experience testing, where designers consistently associate inspiration images with specific concept keywords despite the paper prototype setup and inquire whether the AI system can interpret them accurately. One designer explained their desire for the AI not only to generate the perfect outcome but also to accompany them as they refine their concepts. Designers seek detailed image and text inspiration to interpret the keyword concept and draw further inspiration from the AI outputs.

\subsection{Design outputs}
\subsubsection{Expected and unexpected}
\label{sec:unexpected_expected}
In Question 5 in the workshop, designers were asked to select preferred qualities of the output from the AI generation. The results are shown in Table \ref{tab:output_preference}. Interestingly, designers expressed a desire for both ``expected'' and ``unexpected'' outputs. All designers choose they want unexpected and interesting outputs. All designers indicated a preference for unexpected and interesting outputs, as they value the AI's ability to generate ideas they might not conceive alone. Unexpected outputs from AI could spark new ideas and creative directions.

However, designers also expressed a need for ``expected'' outputs, which align with the image inspiration they provided to the AI. This desire stems from a sense of control over the output. As one designer articulated, design is a process driven by designers, and they want to steer the direction of design. Therefore, the ability of the AI to produce ``expected'' outputs allows designers to use such systems as tools for assistance in their design process.

\begin{table}[h]
\renewcommand{\arraystretch}{1.5}
    \centering
    \begin{tabular}{|p{0.32\textwidth}|c|}
        \hline
         \textbf{Output Preference} & \textbf{Responses}\\
         \hline
         Unexpected and interesting & 6 \\
         \hline
         Representative of the image inspiration &  5\\
         \hline
         Expected &  4\\
         \hline
         Low resolution and rough & 3 \\
         \hline
         Constraint enforced, e.g. symmetric & 2\\
         \hline
         High resolution, clear and sharp & 1 \\
         \hline
         Representative to the text inspiration & 1 \\
         \hline
         Other & 0 \\
         \hline
    \end{tabular}
    \caption{Qualities of design output designers prefer}
    \label{tab:output_preference}
\end{table}

Designers prioritize ideation over the resolution and sharpness of outputs from AI tools. They are content with lower resolution and rough outputs, as they primarily view AI as a tool for generating ideas rather than for detailed design execution. Designers value maintaining fine control over the design outcome, and they harbor skepticism about current AI capabilities to provide the level of detailed control they can achieve themselves.

Furthermore, designers do not anticipate AI to produce outputs that precisely mirror the text inspiration provided, as text serves primarily to set the design direction. They recognize that there can be multiple interpretations of a design idea. Instead, designers perceive image inspiration as playing a more crucial role in controlling the visual aspects of the design.

\subsubsection{Novel and useful} 
\label{sec:novel_useful}
\begin{figure}[h]
    \centering
    \includegraphics[width=0.95\linewidth]{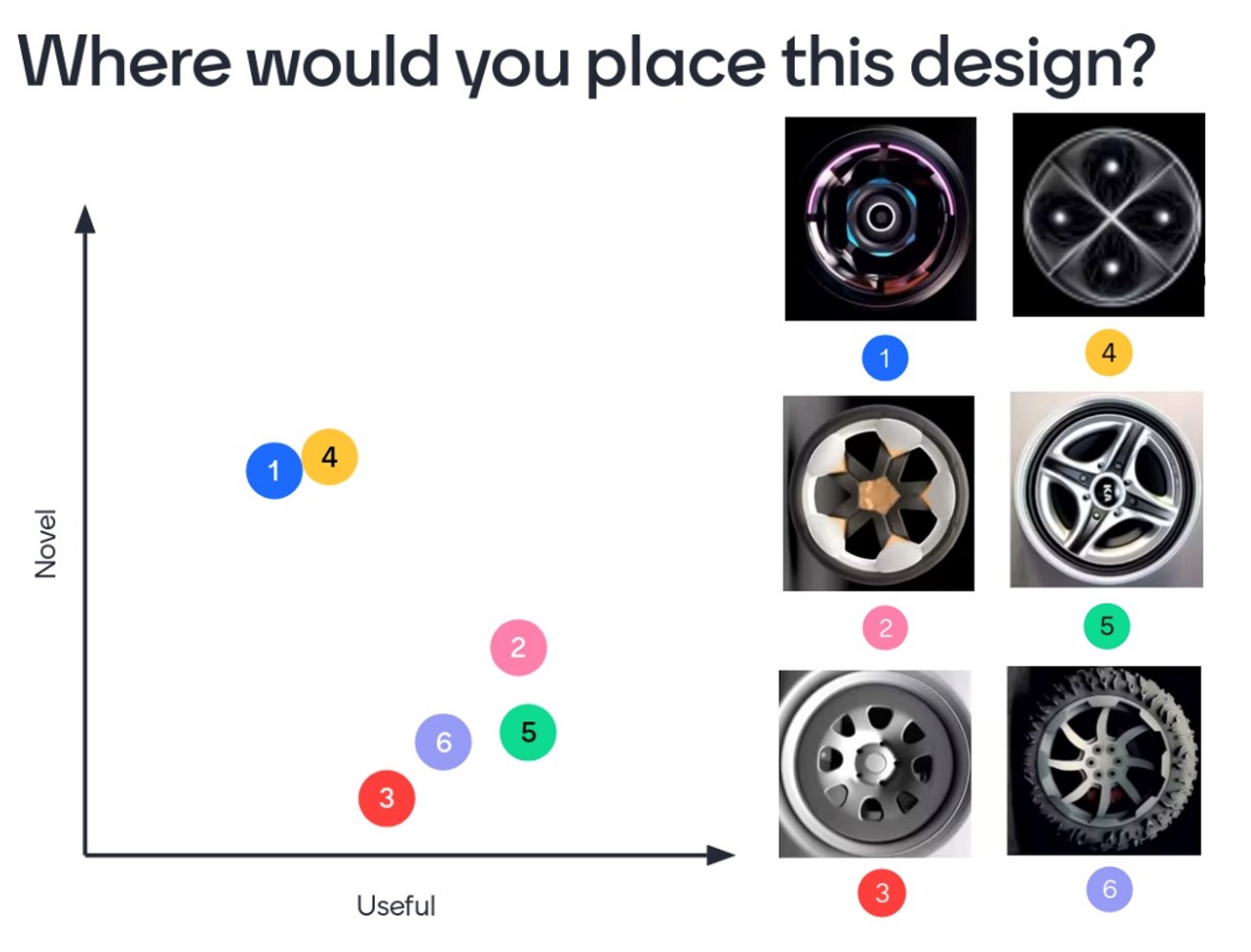}
    \caption{The average scores of novelty (low-high) and usefulness (low-high) that the designers rated 6 designs on during the workshop.}
    \label{fig:novel_and_useful}
\end{figure}

Designers expressed a desire for both "useful" and "novel" designs from the AI. To better understand how these terms correspond to design output, we asked designers to rank six pre-generated designs on usefulness and novelty and discuss their responses (Question 6 during the workshop). The purpose of this exercise was to better understand what kind of designs or design elements designers consider to capture both of these aspects they previously indicated as important metrics. These designs ranged from more functional wheels to more abstract concepts. The results are depicted in Figure \ref{fig:novel_and_useful}.

Novelty in the context of AI generated output appeared to be more closely related to feelings of unexpectedness or surprise, with the specifics of what that looks like depending on the individual designer and their experiences and expectations. For example, the designs 1 and 4 were considered novel but not useful by designers. While they represented new ideas, designers noted that significant additional work would be required to make them practical. Conversely, designs like 3, 5, and 6 were considered potentially useful as wheels, but lacked sufficient novelty to be interesting or inspiring to the designers. Design 2 emerged as an example that represented a new idea and could be feasibly translated into a functional wheel. Throughout the discussion, it became clear that the designers expressed a willingness to sacrifice utility for novelty, as they are confident in their ability to achieve functional designs, but find novelty more engaging and challenging to attain.

\subsection{Design Process Integration}

\subsubsection{Continuous generation}
\label{sec:continuous_generation}
In the workshop poll, designers were prompted in Question 7 to specify when they preferred the generation to occur. They were presented with three options: generating the output only upon clicking ``generate'', generating the output after adding design feedback, and continuously generating in real-time. Designers had the flexibility to choose multiple options. Surprisingly, out of the six designers, five chose continuous real-time generation, one preferred generating the output after adding design feedback, and only one chose generating the output only upon clicking ``generate''.

Designers draw parallels to their experience with image search platforms like Pinterest, where scrolling infinitely is possible. This preference stems from the recognition that inspiration is boundless, and good concepts emerge sporadically. Designers are accustomed to sifting through vast amounts of content to find designs that resonate with them. Continuous generation provides them with an abundance of choices and may ensure that each interaction leads to meaningful directions in their design process.

\subsubsection{Regenerate with feedback}
\label{sec:regenerate_with_feedback}
Subsequently in Question 8, designers were asked to rank the design organization features based on their usefulness. The most useful feature is the ability to provide design feedback on a generated result, followed by assigning a novelty score to a design, star/like a design, assigning a usefulness score, grouping designs, and adding tags to designs.

During the paper prototype testing phase, we observed a consistent inclination towards this preference. Designers directly attached sticky notes as comments to the generated results and requested the system to regenerate them accordingly. Additionally, they exhibited a tendency to swiftly discard designs rather than sorting through them. Designers expressed a desire to seamlessly integrate AI into their creative process, envisioning it as a junior designer aiding in idea generation and collectively critiquing designs to improve them.

\section{Generative AI application for design inspiration}
\label{sec:ai_system}
Drawing from the insights presented in Section \ref{sec:design_inspiration_studies} \nameref{sec:design_inspiration_studies} and Section \ref{sec:interaction_mode} \nameref{sec:interaction_mode}, we developed a functional prototype of a generative AI system aimed at assisting designers with developing concepts using text and image inspiration. User testing was conducted on this prototype, engaging both individual designers and their design teams to gather feedback.

\subsection{User Interface Design}
\begin{figure}[h]
    \centering
    \includegraphics[width=\linewidth]{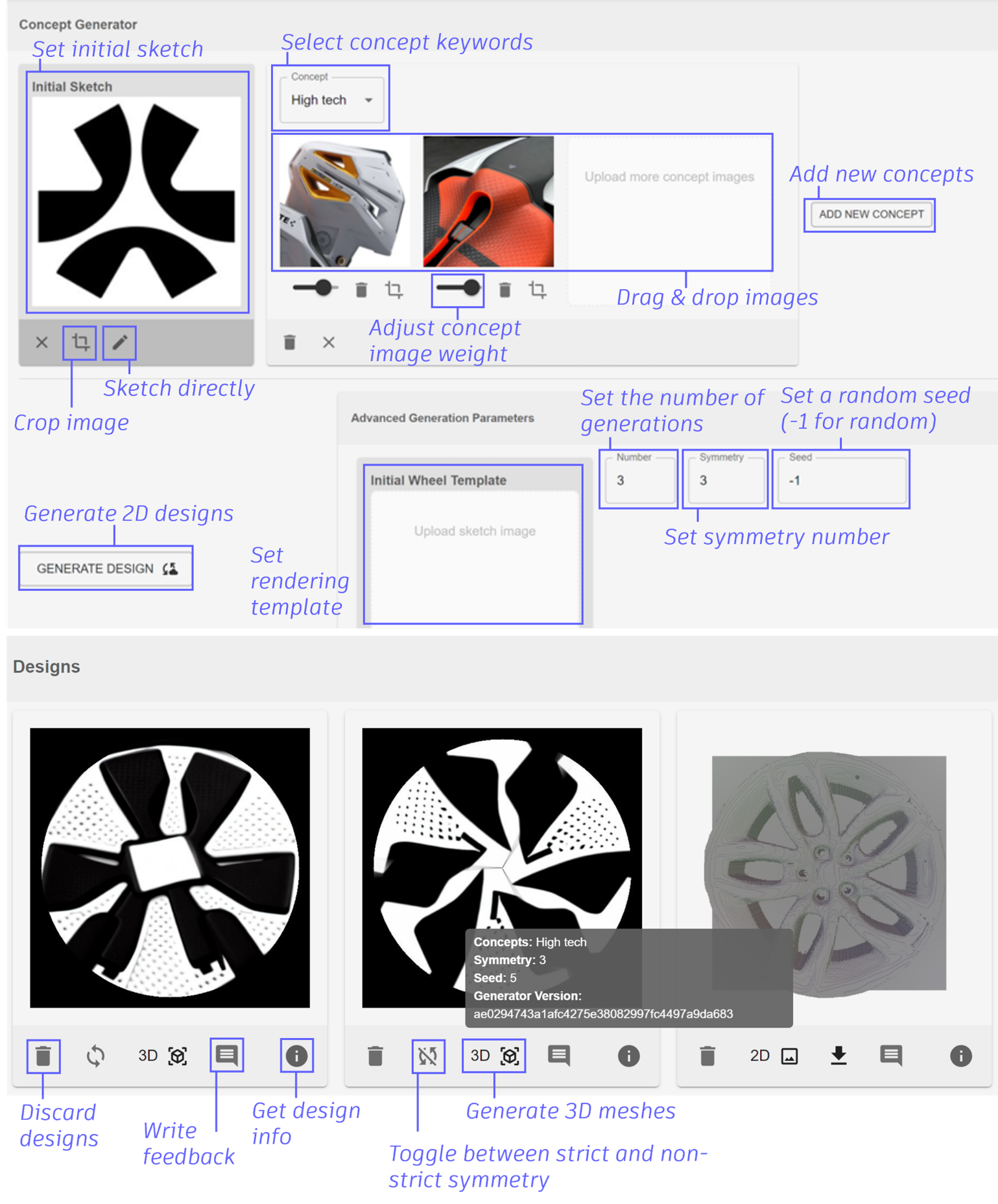}
    \caption{Screenshot of the Generative AI system prototype user interface.}
    \label{fig:interface}
\end{figure}

The screenshot of the user interface is shown in Figure \ref{fig:interface}. Designers start with a rudimentary initial sketch, selecting design keywords to establish design directions. Within each keyword defined direction, they can further refine their concepts by adding image inspirations. Designers also have the flexibility to adjust the symmetry settings to control the repetition in the generated designs and specify the quantity of designs to be generated simultaneously. 

Below, we outline the detailed features prioritized in the AI tool, along with references to corresponding findings in studies on design inspiration and interaction modes that support these priorities:
\begin{itemize}
    \item Applying a \textit{hierarchical approach} on inputting text and image inspiration (Section \ref{sec:hiearchical_approach}). 
    \item Emphasis on using \textit{keywords} as primary text inspiration, focusing on defining concepts and establishing design directions (Section \ref{sec:keyword_direction}, \ref{sec:keyword_meaning} and \ref{sec:keyword_concept_definition}). Other potential uses of keywords, such as shape descriptions and requirements (Section \ref{sec:embody_keywords}), are not addressed in the tool.
    \item Emphasis on using image inspiration to \textit{extract visual elements} (Section \ref{sec:extract_visual_elements}), rather than conveying feelings (Section \ref{sec:convey_feelings}).
    \item Flexibility in the number of inspiration inputted, with testing focused on \textit{1 to 3 keywords} and \textit{1 to 3 image inspiration} for each keyword (Section \ref{sec:quantity_inspiration}). 
    \item Support for generation and regeneration with feedback. Users can select the number of outputs to provide designers with more design options, though continuous generation as desired by designers is not implemented (Section \ref{sec:continuous_generation}). 

\begin{figure}[h]
    \centering
    \includegraphics[width=\linewidth]{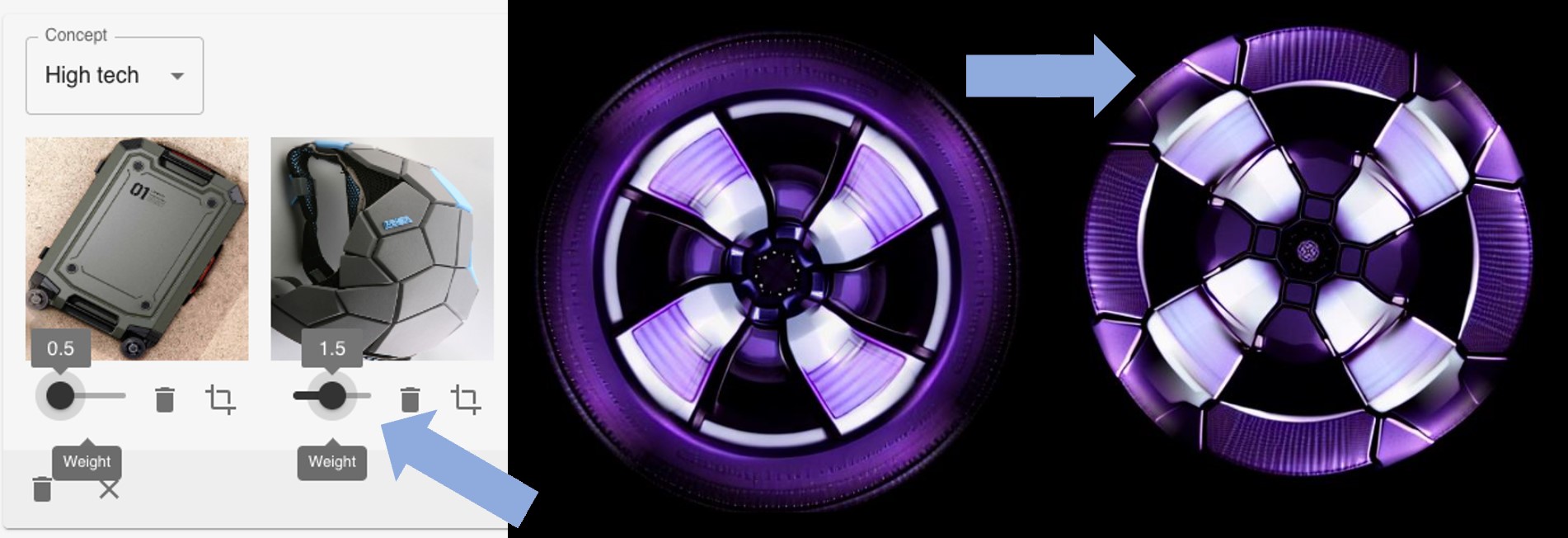}
    \caption{Weight adjustments and the effect on generated designs.}
    \label{fig:weights}
\end{figure}

    \item Support for the generation of \textit{expected} designs (Section \ref{sec:unexpected_expected}) with features such as image cropping and inspiration weight adjustments to grant designers greater control over the generation process.   

\begin{figure}[h]
    \centering
    \includegraphics[width=\linewidth]{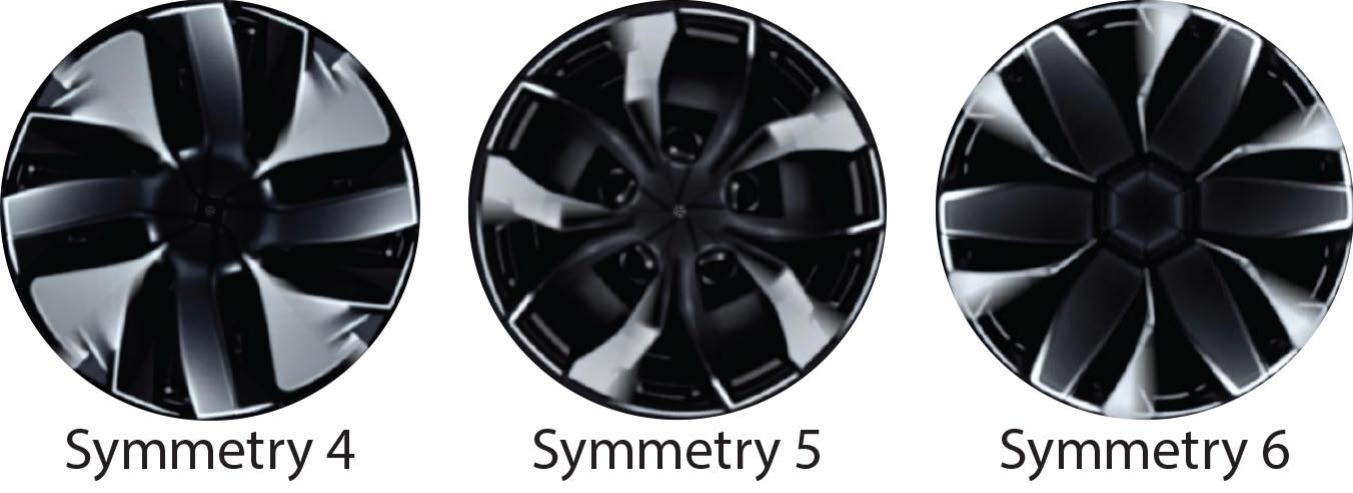}
    \caption{Designs with same inspiration but different number of symmetry repetition.}
    \label{fig:symmetry_number}
\end{figure}

\begin{figure}[h]
    \centering
    \includegraphics[width=\linewidth]{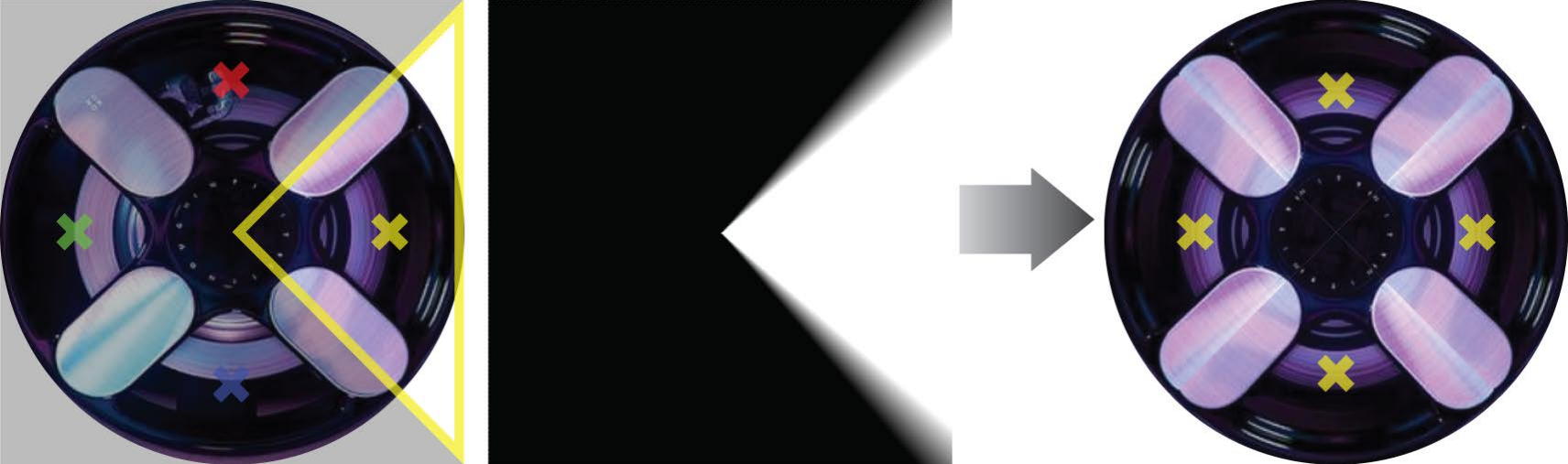}
    \caption{Symmetry constraint.}
    \label{fig:symmetry}
\end{figure}

    \item Support for the generation of  \textit{useful} designs (Section \ref{sec:novel_useful}). Designers can input symmetry repetition numbers (Figure \ref{fig:symmetry_number}), and symmetry constraints are integrated into the AI model to generate symmetric wheel designs (Figure \ref{fig:symmetry}).
    \item Additionally, designers have the option to toggle off symmetry to facilitate the generation of more \textit{unexpected} designs (Section \ref{sec:novel_useful}). 
\end{itemize}

\subsection{AI Model Architecture}
\label{sec:ai_model_architecture}

\begin{figure*}[ht]
    \centering
    \includegraphics[width=0.9\linewidth]{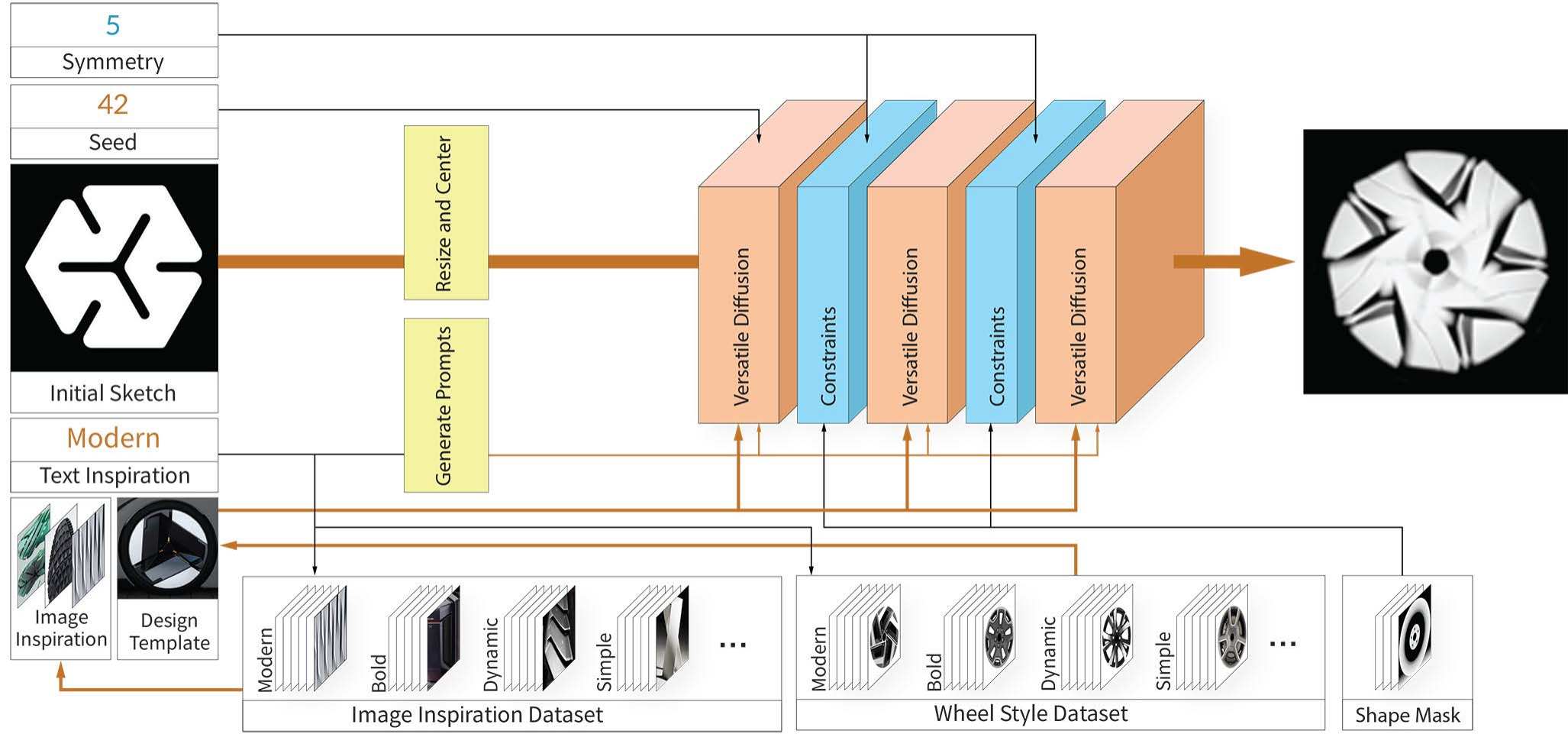}
    \caption{Architecture of the AI model, user inputs, and the image datasets that are optionally used to provide initial sketch, wheel template, and/or image inspirations at the generation time.}
    \label{fig:arch}
\end{figure*}

The AI model for concept generation, based on both image and text inspiration, is required to integrate visual elements from image inspiration and understand the stylistic meaning behind text inspiration. Additionally, the model needs to generate designs that exhibit key characteristics of wheels, including their general shape, position, and orientation.

To address this requirement, we evaluated four different diffusion models. Section \ref{sec:generative_ai} \nameref{sec:generative_ai} introduces the backgrounds on diffusion models and explains why they are suitable for design tasks.

\begin{itemize}
\item \textit{Stable Diffusion} \cite{rombach2022high} is a commonly used model, but it only accepts a textual prompt as input and cannot incorporate image inspiration.
\item \textit{ControlNet} \cite{zhang2023adding} and its successors allow more precise control of the general shape, position, and orientation of the wheel through different conditional inputs such as depth and normal maps, segmentation maps, and line drawings. However, they do not provide stylistic control beyond a color palette.
\item \textit{Kandinsky} \cite{razzhigaev2023kandinsky} can support both image and text inspiration by mapping multiple image and text inputs into a shared latent space and mixing them together. However, the semantics and functionality of the generated images are affected by the image inspirations and may not be preserved. For example, using an image of a helmet as inspiration could result in the wheel design resembling a ball or a helmet instead of incorporating stylistic elements from the helmet.
\item \textit{Versatile Diffusion} \cite{xu2023versatile} allows a textual prompt and multiple images to influence the generated images at the semantic or stylistic level using an optional global control signal and cross-attention mechanisms. However, it lacks fine-grained control inputs (similar to ControlNet) or an in-painting model, making it challenging to control the general shape, position, and orientation of the generated wheel.
\end{itemize}
We chose \textit{Versatile Diffusion} as our base model for its flexibility in incorporating both image and text inputs to influence generation at the semantic and stylistic levels. Built upon this foundation, our AI model is shown in Figure \ref{fig:arch}. The initial sketch from the user is used with some added noise as the initial input to the diffusion model. The wheel template image from the user or the image inspiration dataset is fed into the versatile diffusion's cross-attention layer through a global semantic route, while the image inspirations from the user or sampled from the dataset are fed into the cross-attention layers through a non-global route \cite{xu2023versatile}. After a number of denoising steps in the first denoising sub-process, the radial symmetry constraint and masking are applied to the generated image, and the result is fed into the next denoising sub-process. 

This architecture addresses two challenges: 
\begin{enumerate}
    \item Open-source generative models are not trained with enough stylistic keywords and, therefore, are not sensitive to such cues in the prompt.
    \item  Radial symmetry constraints must be imposed on the generated wheel.
\end{enumerate}

To tackle the first challenge, the keywords are added to a default prompt to influence generation. However, since the model may not be sensitive enough to stylistic keywords when the user does not provide any image inspiration, the stylistic keywords are also used to sample images from two image datasets. Images of wheel designs with the given stylistic label are randomly sampled from a dataset of wheel designs to be used as the initial sketch and/or the wheel template if the user does not provide any. Moreover, if the user does not provide any image inspirations, up to three stylistic images (labeled with the given keyword) are similarly sampled from a second dataset of image inspirations.

To address second challenge of radial symmetry, we divide the diffusion model's denoising process into multiple sub-processes and impose the symmetry, position, and orientation constraints for the wheel in between. Masking and symmetrical replication are used between the sub-processes, and an optional symmetrical replication step is employed at the end to enforce the constraints. Figure \ref{fig:symmetry} illustrates an example of a mask used in the replication process for imposing the radial symmetry constraint (for a symmetry number of 4). A circular mask is also utilized to enforce the shape, position, and orientation constraints. A final masking and an optional symmetrical replication are applied at the end to generate the final wheel design.

Better methods of imposing radial symmetry and general shape, position, and orientation constraints, such as in-painting, can alternatively be used in the future. Additionally, fine-tuning the language model used for the input prompt to better capture stylistic keywords with improved or curated datasets can be explored.

\subsection{Application Architecture}
The application's architecture is deployed on Amazon Web Services (AWS), utilizing GPU-accelerated EC2 instances (p3.2xlarge) with NVIDIA V100 GPUs, each providing 16GB of video RAM. These instances were chosen to meet high-performance computing needs essential for executing complex operations efficiently. To ensure scalability and effective resource management, EC2 Auto Scaling groups are implemented, allowing the system to adjust computational resources dynamically in response to changing workloads.

The front-end of the application is constructed using TypeScript, React, and Next.js, incorporating Autodesk Platform Service for authentication. This setup is based on the T3 stack, indicating a selection of technologies known for their robustness and developer-friendly nature in creating scalable web applications. The front-end serves as the user interface, facilitating secure access and interaction with the application's services. Among these services is the diffusion wheel design generator, which leverages theme images and initial designs to generate customized wheel designs, showcasing the application's capability to produce tailored outputs based on specific inputs.

\subsection{User Feedback}
\label{sec:feedback}
We conducted three 30-minute testing sessions with individual designers. The examples generated by designers in these sessions are shown in Figure \ref{fig:design_example}. Following this, we organized a 90-minute in-person group design session involving twelve designers. In this session, one designer controlled the application while sharing the screen on a large projector, while others provided live suggestions on inspiration and actions within the application. Designers were encouraged to think aloud in both testing scenarios.

Overall, designers found the interface highly intuitive. They were able to use all features with minimal guidance. Designers used Pinterest to drag images for inspiration alongside their initial sketches into our AI application. They selected three to six images for simultaneous generation. Designers particularly appreciated the ability to reinforce symmetry, often adjusting the symmetry number to explore various design variations and save time. 

They also liked the rapid incorporation of visual elements from inspiration images into designs, enabling them to start with simple inspiration and explore a large quantity of generated and unique designs. The cropping feature was well-received for its ability to focus on specific parts of an image, and designers experimented with weights on every generation to better understand the AI's generation.

However, designers expressed a desire for finer control over the design generation process. While they found designs they liked, they felt these were often incidental rather than deliberately crafted. Suggestions for improvement included masking and the ability to extract detailed visual features identified by the AI for enhanced control and refinement.

Furthermore, designers sought more explanation about the generated results, both for reproducibility and to understand how the AI generated them. They desired AI to incorporate their feedback into subsequent generations of designs, emphasizing the need for rationale behind design directions inspired by generated designs in order to integrate such AI tools effectively into their design processes.

\section{Limitations and Future Work}
This work provides a unique opportunity to contextualize the challenges in conceptual automotive design and test the limits of cutting-edge generative AI research. We developed the final prototype and tested it with designers. This closed-loop approach allowed us to validate our assumptions and identify future opportunities. The most important feedback regarding the tool itself was that while the designers found the interaction with our generative AI tool to be intuitive, they desired more control over the design outcomes and more explainable results. These functional limitations are detailed in Section \ref{sec:feedback} \nameref{sec:feedback}.

Centering this work on one company allowed us to propose novel and meaningful interactions with generative AI in an automotive context. The automotive industry is undergoing significant changes with the rise of startups and the growing demand for electric vehicles. This shift requires innovative design approaches due to the fundamental differences in power provision and consumer expectations. The competitive market demands faster development cycles, increasing the pressure on generating and executing innovative ideas. Integrated design processes, where design and engineering teams collaborate closely, are becoming more common. Although this paper does not explore these aspects, we see significant importance in researching generative AI interactions grounded in real-world design processes with actual designers and engineers to ensure responsible and meaningful outcomes. In particular, future research should explore generative AI interactions across various organizations and industries to map the boundaries and applicability of generative AI tools.

\section{Conclusion}
The rapid evolution of generative AI technology significantly impacts the design domain, particularly within conceptual design, where its open-ended nature and reliance on creative ideas provide numerous opportunities for AI creativity to enhance human creativity. This paper seeks to explore three fundamental questions: 1. How do designers currently use inspiration in their conceptual design workflows? 2. What preferences do designers have regarding interaction with generative AI systems for concept development? 3. How can a generative AI application be developed to meet these identified needs and design practices?

Despite the widespread availability of generative AI technology, its meaningful integration into design processes remains constrained. Understanding that designs happen both individually and collaboratively, and that a wealth of knowledge is embedded and shared within a common design context, we employed a diverse set of methods—including surveys, workshops, and user testing—and recruited designers and design teams that share similar design contexts. This integrated approach allowed us to investigate the three research questions in depth without sacrificing generality, and enabled end-to-end testing of the final development with the intended end users.

Technologies for design must strike a balance between harnessing human creativity and adhering to human-centered design practices. Beyond demonstrating the powerful capabilities of AI in assisting designers in conceptual design, our goal is to spark thoughtful inquiry into how to develop future AI systems for design responsively and the potential impact of these systems on future design practices.


\section{Acknowledgement}
The authors extend our gratitude to Brandon Cramer, Pradeep Kumar Jayaraman, and Aditya Sanghi for their valuable advice on the project, and to Ally Park, Jiwon Jun, and Alberto Tono for their facilitation of the studies. 
Special thanks to Christine Toh 
her insightful suggestions on the manuscript. 
Lastly, appreciation goes to all the designers from Hyundai Motor Company who participated in the studies for generously sharing their design expertise.

\bibliographystyle{asmeconf}  
\bibliography{references}

\end{document}